\documentclass[%
 aip,
 jap,
 amsmath,amssymb,
 reprint,
]{revtex4-1}

\usepackage{graphicx}
\usepackage[caption = false]{subfig}
\usepackage{dcolumn}
\usepackage{bm}

\usepackage[utf8]{inputenc}
\usepackage[T1]{fontenc}
\usepackage{mathptmx}
\usepackage{xcolor}

\begin{document}

\preprint{AIP/123-QED}

\title{Optimal Filtering of Overlapped Pulses in Microcalorimeter Data}

\author{Dallas Wulf}
\email[]{dallas.wulf@mcgill.ca}
\altaffiliation{Current Address: McGill University, 3600 rue University, Montreal, QC H3A 2T8, Canada}
\affiliation{University of Wisconsin-Madison, 1150 University Avenue, Madison, WI 53706, USA}
\author{Felix Jaeckel}
\affiliation{University of Wisconsin-Madison, 1150 University Avenue, Madison, WI 53706, USA}
\author{Dan McCammon}
\affiliation{University of Wisconsin-Madison, 1150 University Avenue, Madison, WI 53706, USA}
\author{James A Chervenak}
\affiliation{NASA/Goddard Space Flight Center, 8800 Greenbelt Road, Greenbelt, MD 20771, USA}
\author{Megan E Eckart}
\altaffiliation{Current Address: Lawrence Livermore National Laboratory, 7000 East Avenue, Livermore, CA 94550, USA}
\affiliation{NASA/Goddard Space Flight Center, 8800 Greenbelt Road, Greenbelt, MD 20771, USA}

\date{\today}

\begin{abstract}
Here we present a general algorithm for processing microcalorimeter data with special applicability to data with high photon count rates. Conventional optimal filtering, which has become ubiquitous in microcalorimeter data processing, suffers from its inability to recover overlapped pulses without sacrificing spectral resolution. The technique presented here was developed to address this particular shortcoming, and does so without imposing any assumptions beyond those made by the conventional technique. We demonstrate the algorithm's performance with a data set that approximately satisfies these assumptions, and which is representative of a wide range of microcalorimeter applications. We also apply the technique to a highly non-linear data set, examining the impact on performance in the limit that these assumptions break down.
\end{abstract}

\maketitle

\section{Introduction}\label{sec:intro}

Optimal filtering is a well established technique for processing data from microcalorimeter detectors. It assumes that all photons absorbed by the detector produce a pulse with a common signal shape $S(t)$, scaled by an amplitude that is proportional to the incident photon energy. To recover the photon energy with optimum resolution, this technique constructs the pulse amplitude estimator 

\begin{equation}
\label{eq:estim}
E = \sum_{j=1}^{\infty} w_{j} s_{j}
\end{equation}

\noindent
where $s_{j}$ is the measured signal amplitude in the $j^{th}$ frequency bin, equivalent to the discrete Fourier transform of $S(t)$. The weights $w_{j}$ are chosen to maximize the signal-to-noise ratio of $S(t)$ relative to the detector noise. Under the further assumption that the statistical properties of the noise are not changing with time, it can readily be shown that the weights $w_{j}$ are given by 

\begin{equation}
\label{eq:weights}
w_{j}=s^{*}_{j}/n^{2}_{j}
\end{equation}

\noindent
where $s^{*}_{j}$ and $n^{2}_{j}$ are the complex conjugate of the mean signal and the mean-square of the noise in the $j^{th}$ frequency bin, respectively (for a complete derivation, see e.g. Ref.~\onlinecite{szymkowiak_signal_1993}). Back in the time domain, this process is equivalent to the convolution of the signal $S(t)$ with the inverse Fourier Transform of $w_{j}$, $W(t)$. Whereas each time sample of the unfiltered signal $S(t)$ provides correlated information about the pulse amplitude $E$, the best estimate of this quantity is given by just a single point in the filtered signal, i.e.

\begin{equation}
\label{eq:convo}
E = W(t) \otimes S(t)|_{t=0}
\end{equation}

\noindent where $t=0$ corresponds to the pulse arrival time.

In practice, the detector output is sampled and digitized at a rate sufficient to avoid aliasing the useful terms in Eq.~\ref{eq:estim}. The useful bandwidth extends up to frequency $f_{max}$, beyond which additional terms will not significantly improve the signal to noise ratio. The time-domain length of $W(t)$ determines the frequency resolution of the filter, with longer filters having finer frequency-space binning. Longer filters are thus more efficient for optimizing the signal-to-noise ratio (particularly when the noise spectrum contains features), providing improved energy resolution. To achieve the best possible energy resolution, this length must typically be many times the decay time of the pulse, $\tau$.

\begin{figure}
\includegraphics[width=3.37in]{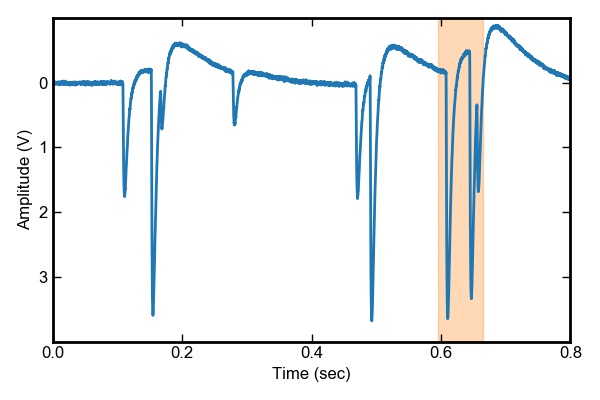}
\caption{Example of pulse overlap in XQC data ($\tau = 9$ ms). Each of the nine pulses in this data segment have a minimum separation of less than the 200 ms time-domain filter length typically used for this detector for best energy resolution, and therefore cannot be processed using COF techniques. Using the OLPF technique presented here, however, each of these pulses may be recovered without sacrificing resolution. The shaded region indicates the time interval of the filtered output shown in Fig.~\ref{fig:filDat}.}
\label{fig:pileup}
\end{figure}

One major drawback of conventional optimal filtering (henceforth COF) is its requirement that pulses produced by individual photons have a minimum separation of at least the time-domain length of the optimized filter, $W(t)$. Pulses with less than this minimum separation cannot be processed, resulting in detector dead-time (Fig.~\ref{fig:pileup}). Practical considerations on detector design place a limit on detector speed (i.e. minimum $\tau$), meaning that $W(t)$ cannot be shortened arbitrarily without sacrificing performance. Consequently, a compromise must usually be made between the desired spectral resolution and the dead-time when choosing the filter length. 

In response to this problem, we have developed a technique called ``Overlapped Pulse Fitting'' (OLPF) for recovering overlapped pulses without loss of resolution compared to COF. Early implementations of this technique were tailored to individual, relatively small data sets and required significant human intervention (Refs.~\onlinecite{crowder_observed_2012,wulf_technique_2016}). The current algorithm has been generalized to be applicable to all microcalorimeter data sets that approximately satisfy the two assumptions made by COF, namely linear response and stationary noise. In addition to the two data sets presented in this paper, this technique was also used to analyze three sounding rocket observations of the soft X-ray background presented in Ref.~\onlinecite{wulf_high_2019}. A different approach to multi-pulse fitting has been presented in Ref.~\onlinecite{fowler_microcalorimeter_2015}, which trades some loss of resolution for computational simplicity.

The OLPF technique was largely developed using data from the University of Wisconsin-Madison/Goddard Space Flight Center X-ray Quantum Calorimeter sounding rocket payload (XQC). The reader may refer to Ref.~\onlinecite{mccammon_high_2002} for a complete description of the payload. XQC is a 36-pixel, silicon thermistor microcalorimeter detector optimized for X-ray energies below 1~keV. Most pixels achieve a baseline resolution of $\sim$7~eV~FWHM under typical laboratory noise conditions. The pixels have an effective decay time $\tau \sim 9$~ms, and are about $3\%$ non-linear at 3~keV. Each pixel is approximately current biased, and the pixel voltages are digitized to 12~bits at a sampling rate of 10.4~kSamples/s.

\section{Technique}

\subsection{Filtering}\label{sec:filt}

\begin{figure*}
\subfloat[\label{fig:avgPls}Average Pulse Shape]{
\includegraphics[width=3.37in]{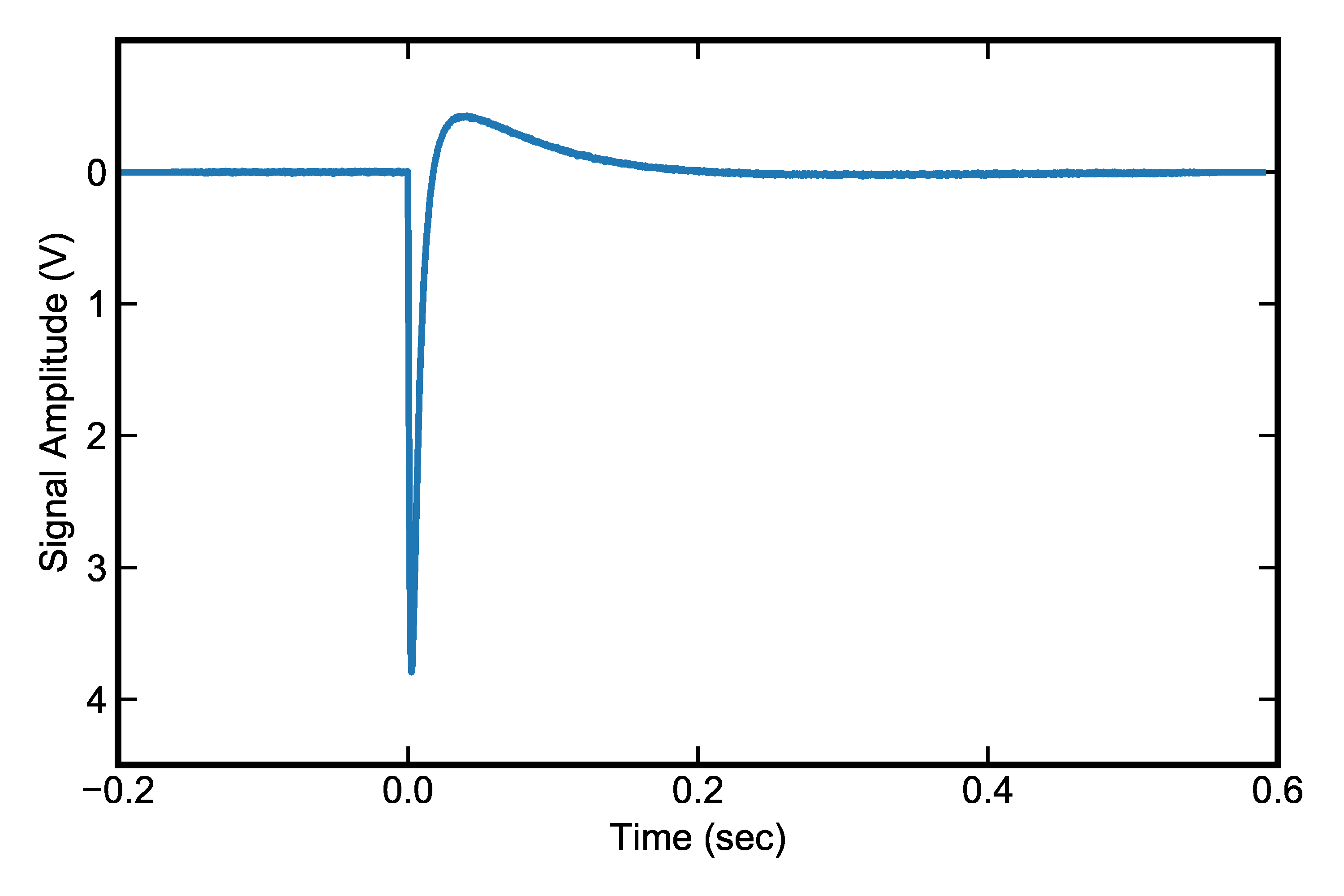}}
\subfloat[\label{fig:avgNos}Average Noise]{
\includegraphics[width=3.37in]{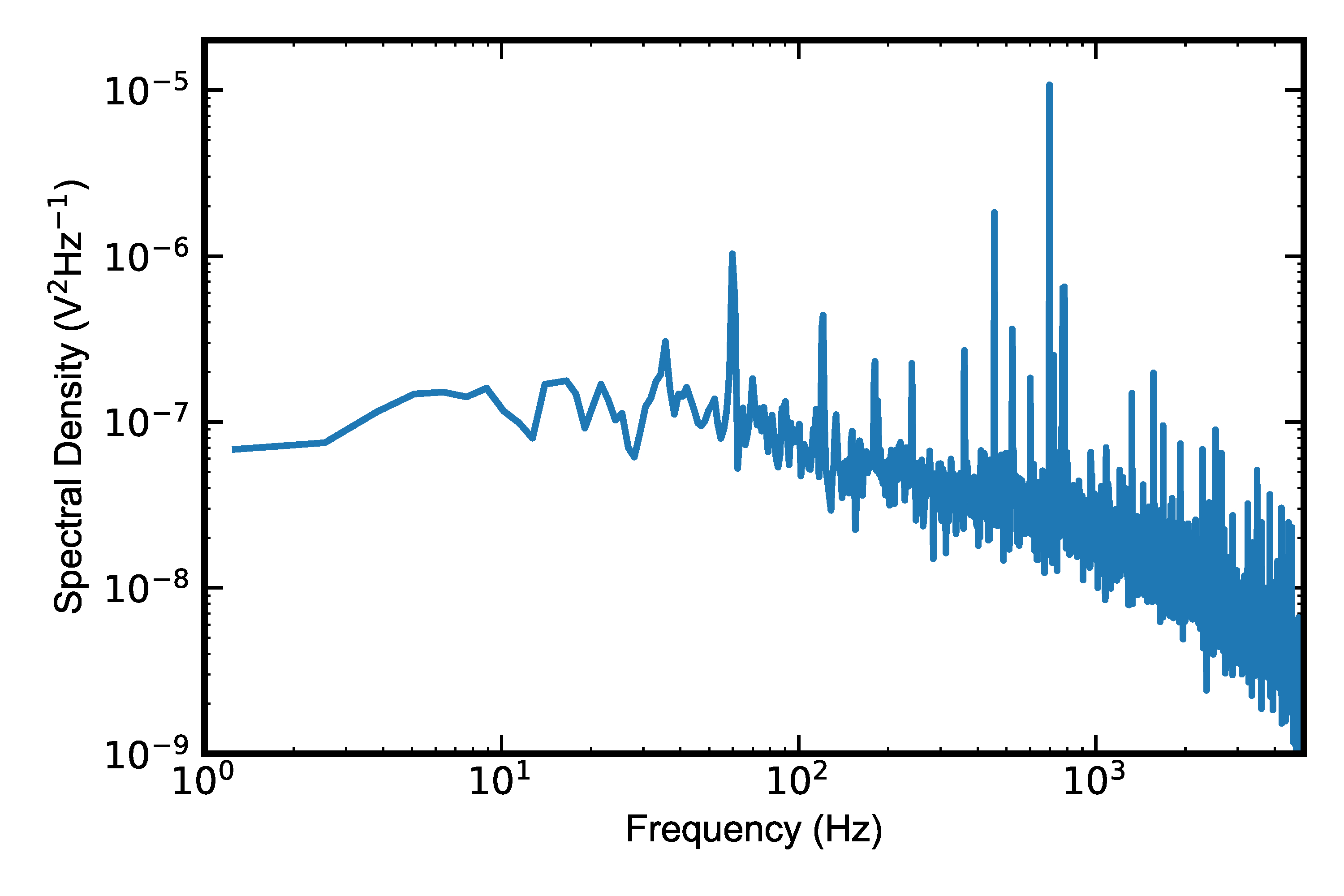}}
\hspace{0mm}
\subfloat[\label{fig:phPls}Pulse Filtered for Amplitude]{
\includegraphics[width=3.37in]{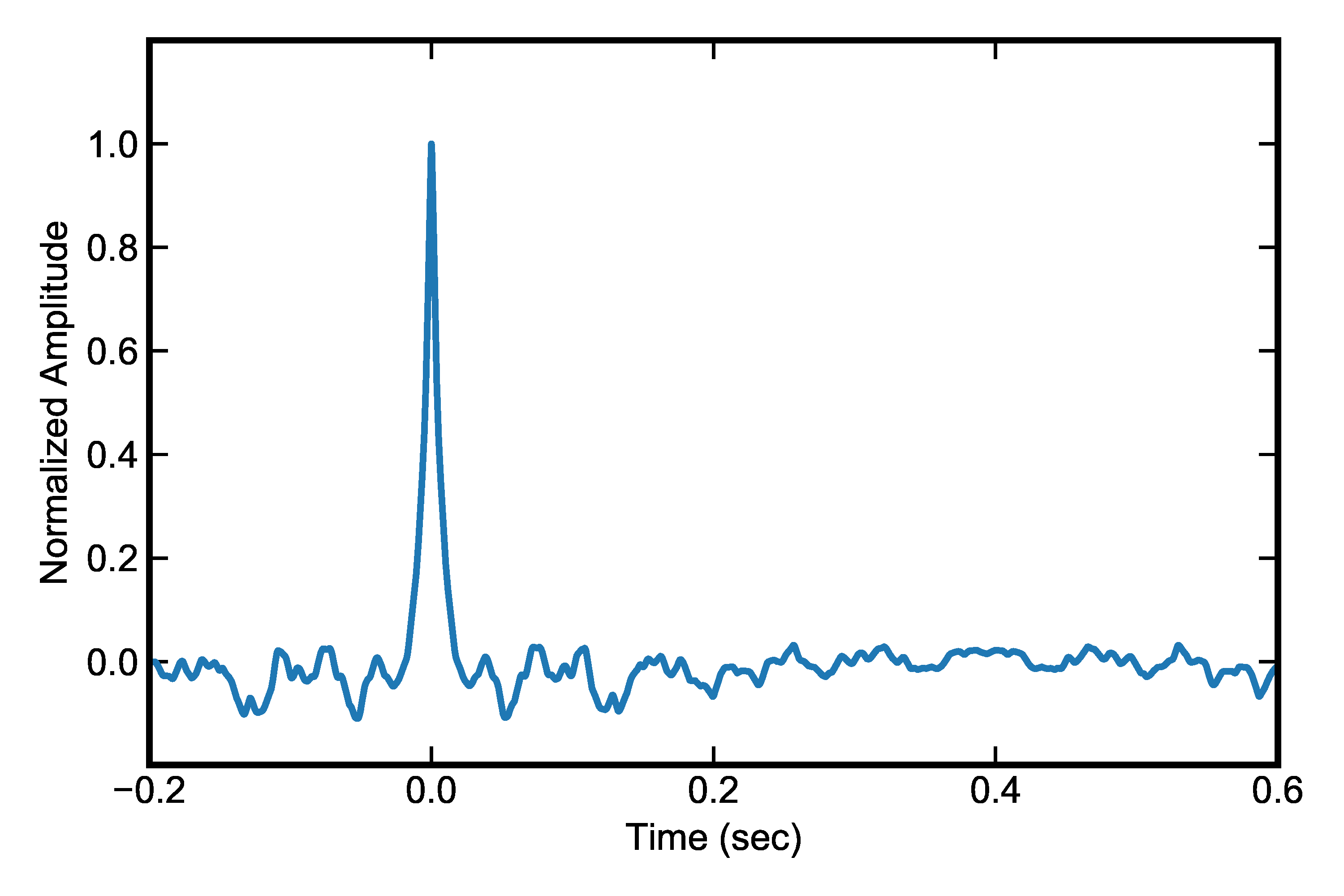}}
\subfloat[\label{fig:ptPls}Pulse Filtered for Arrival Time]{
\includegraphics[width=3.37in]{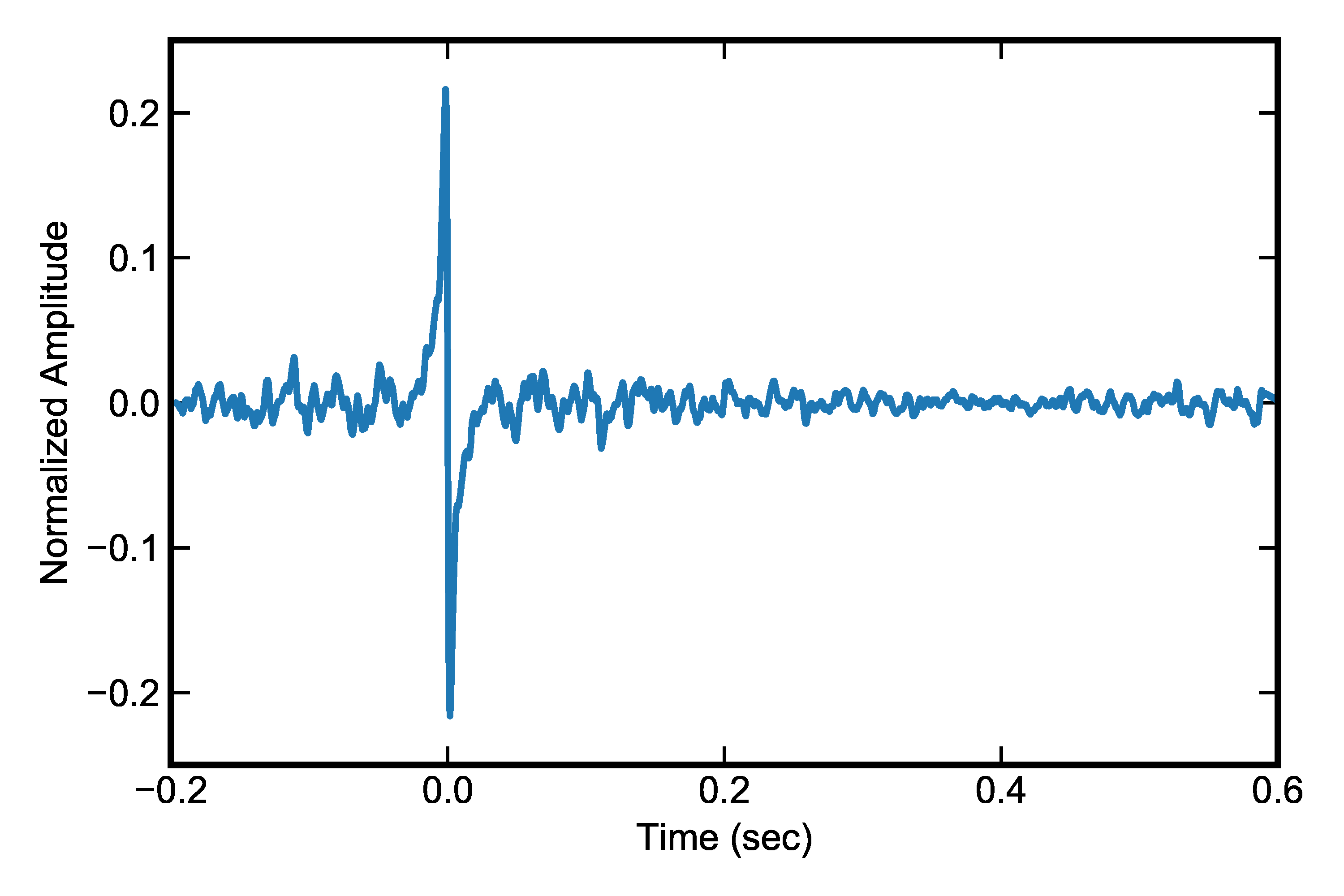}}
\caption{(a) Average pulse template made by averaging several isolated 3.3 keV calibration pulses. (b) Average detector noise spectral density, also made from calibration data. (c) Average pulse template from Fig.~\ref{fig:avgPls} after being optimally filtered for amplitude and peak normalized. As in COF, the value at the pulse arrival time gives the best (highest signal-to-noise) estimate of pulse amplitude. (d) Average pulse template from Fig.~\ref{fig:avgPls} after being optimally filtered for arrival time and normalized such that the mean-square fluctuations of the filtered noise are the same as the amplitude-filtered data. The zero-crossing of this template gives the best estimate of pulse arrival time. In both filtered pulse templates the fluctuations around zero amplitude are not due to noise but are dominated by ringing introduced by the filter, and are therefore exactly predictable for a given pulse amplitude and arrival time.}
\label{fig:pulses}
\end{figure*}

To begin, we construct the usual filter optimized for pulse amplitude described by Eq.~\ref{eq:weights}. In practice, the $s_{j}$ and $n_{j}$ are determined by averaging pulses and noise spectra from calibration data (Figs.~\ref{fig:avgPls}--\ref{fig:avgNos}). Since the dead-time of OLPF does not depend on filter length, this length can be as long as necessary to maximize the spectral resolution, and/or as long as is necessary to contain the entire signal shape. The latter requirement ensures that the signal is zero at the endpoints and the signal shape is not meaningfully altered by windowing. 

In addition to the usual amplitude filter, a filter optimized for pulse arrival time is also constructed. This filter is needed to constrain the pulse's inter-sample arrival time, since sub-sample shifts in time can affect the apparent signal shape and therefore the reconstructed energy of the pulse. To understand the derivation of this filter, it is instructive to revisit the amplitude filter in slightly more detail. The complex phase of the weights in Eq.~\ref{eq:weights} are chosen to make each term in Eq.~\ref{eq:estim} entirely real, and thus maximize the sum by ensuring constructive phase addition of each term. Another consequence of this choice is that the time-domain convolution of the filter with the signal shape can be expressed as a pure cosine sum, with each term constructively adding to attain a global maximum equal to $E$ at time $t=0$ (Eq.~\ref{eq:convo}). Therefore, the output of the amplitude filter for an isolated pulse will be maximum at the pulse arrival time. We can take advantage of this fact to construct an arrival time filter that maximizes the slope of the signal relative to the noise and whose time-domain convolution with the signal shape will have a zero-crossing at $t=0$ (i.e. the pulse arrival time). Rather than deriving such a filter from scratch, we can invoke the Derivative Theorem for Fourier Transforms, simply multiplying the amplitude filter a factor of $2 \pi i f$. Based on the construction of Eq.~\ref{eq:estim}, the corresponding estimator for the slope can be written as

\begin{equation}
\label{eq:deriv}
E' = \sum_{j=1}^{\infty} 2 \pi i f_{j}w_{j} s_{j} = \sum_{j=1}^{\infty} w'_{j} s_{j}
\end{equation}

\noindent
where the $w_{j}$ are the same as in Eq.~\ref{eq:weights}. Here $f_{j}$ is the central frequency of bin $j$, and $i$ is the complex number $\sqrt{-1}$.

The next step is to convolve each of these filters separately with a copy of the entire pixel data stream. This is in contrast to COF, where the amplitude filter is convolved with short, individually triggered data segments containing a single pulse. In practice, these large continuous convolutions can be done efficiently in the frequency-domain and account for only a small fraction of the total computational cost. For ease of processing, these two filtered data streams are split into short segments typically a few times the length of the filter and containing several pulses. Finally, templates for fitting pulses in the filtered data streams are made by convolving the average signal shape with the respective filters (Figs.~\ref{fig:phPls}--\ref{fig:ptPls}). Because pulses in the data stream arrive randomly with respect to the sampling times, a dictionary of fitting templates are precomputed for (typically 10) discreet sub-sample arrival times. Provided that all of the signal power is below the Nyquist frequency, an exact interpolation can be done by advancing the phases in frequency space, via the Shift Theorem for Fourier Transforms.

\subsection{Fitting}\label{sec:fit}

\begin{figure}
\subfloat[\label{fig:phPilup}Amplitude Filtered Data]{
\includegraphics[width=3.37in]{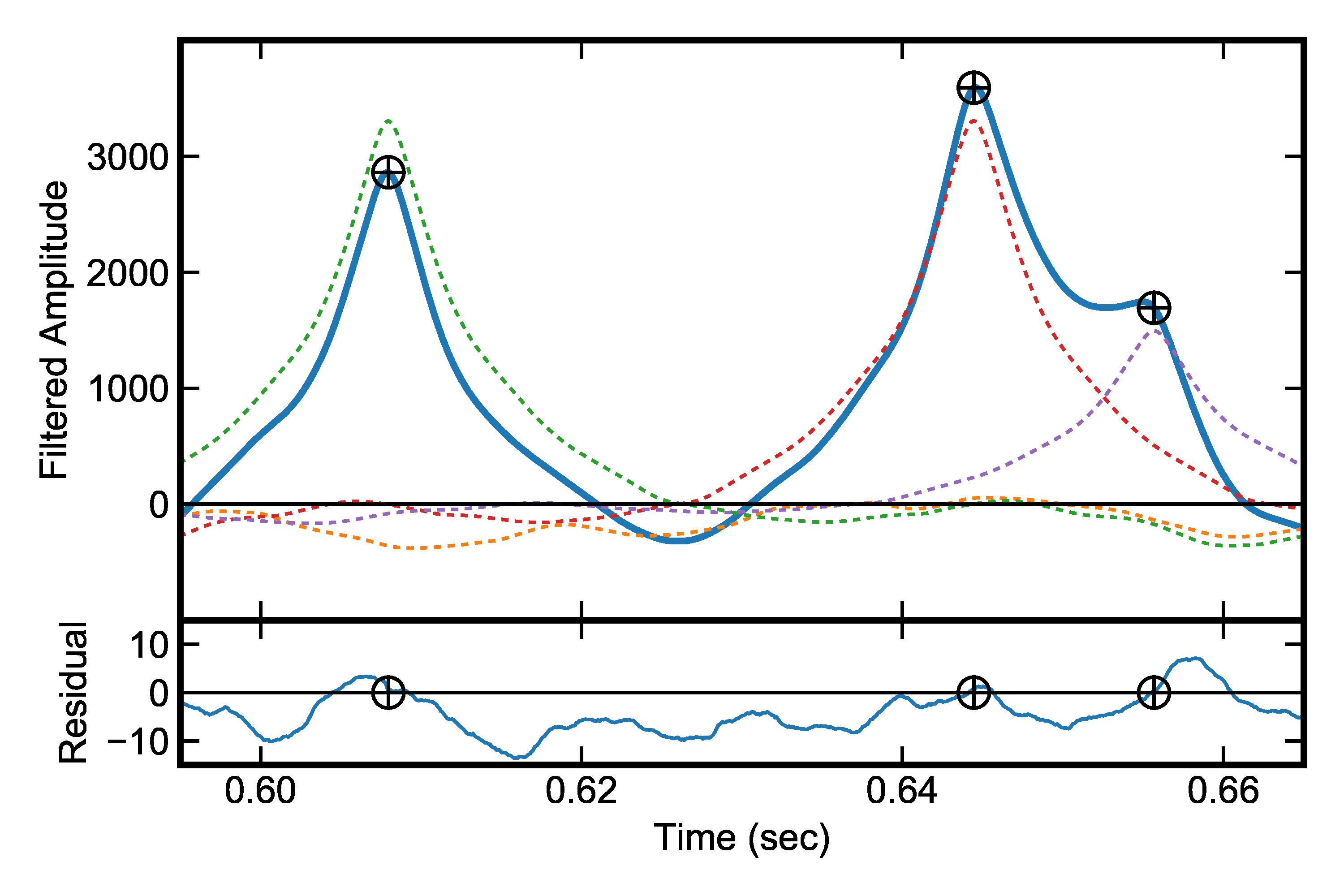}}
\hspace{0mm}
\subfloat[\label{fig:ptPilup}Arrival Time Filtered Data]{
\includegraphics[width=3.37in]{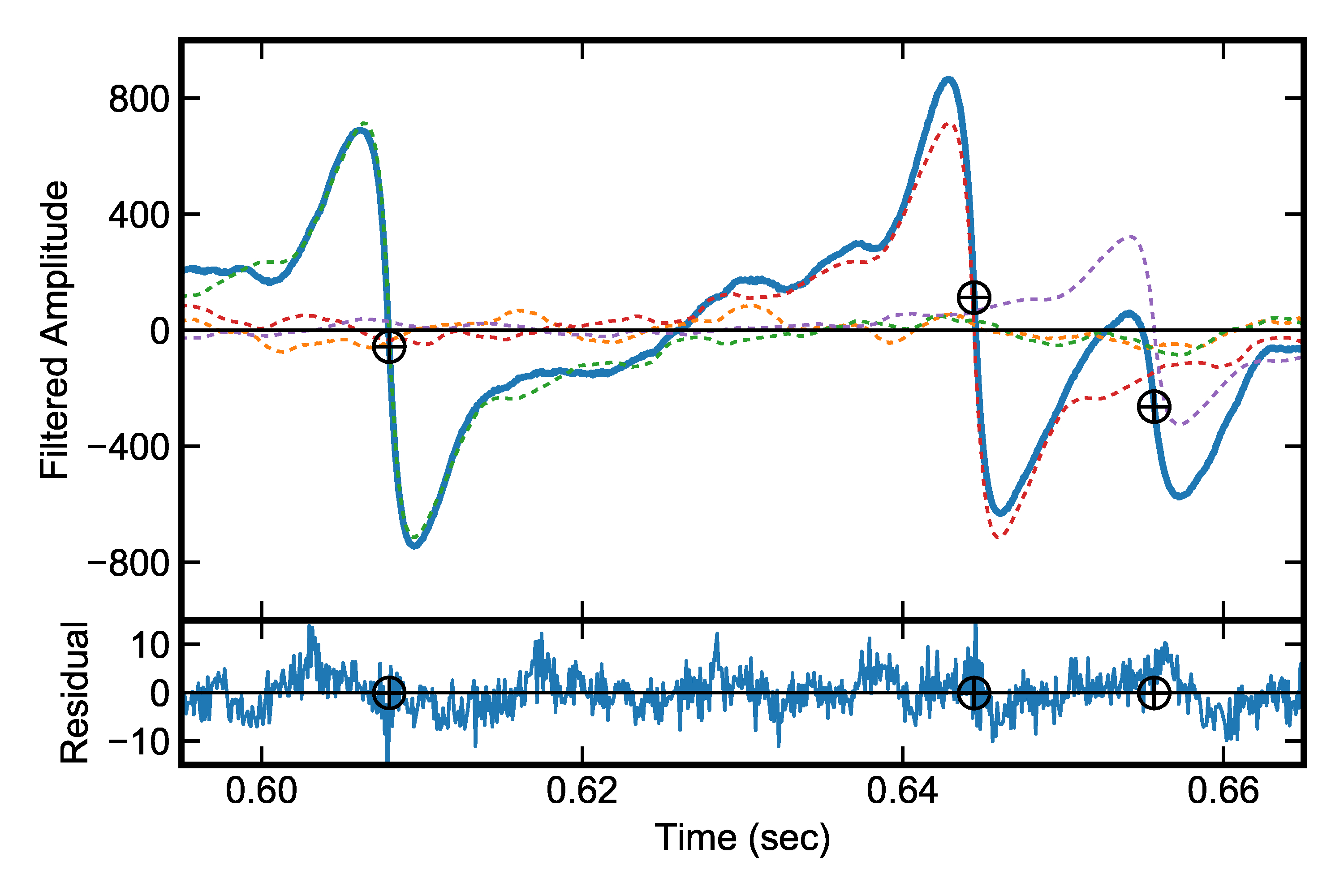}}
\caption[Filtered Pileup]{Data segment from the shaded region of Fig.~\ref{fig:pileup} after optimally filtering for (a) amplitude and (b) arrival time. Filtered data is shown with a solid line, and the model components for each pulse (i.e. the terms in Eq.~\ref{eq:model}) are shown with dotted lines. Crossed circles mark the total model (i.e sum of all components) at the pulse arrival times, and are the only points considered in the solution. The difference between the data and model is shown for all times in the residual panel of each plot. Note that the residuals are minimized at the pulse arrival times.}
\label{fig:filDat}
\end{figure}

For an isolated pulse, the zero-crossing time in the arrival time-filtered data and the value of the amplitude-filtered data at this time give the best estimates of the pulse arrival time and amplitude, respectively. There is no additional information in any of the other points. Things are more complicated for pulses closely spaced in time, since the tail of a pulse and ringing introduced by the filter distorts the signal of every other pulse within one filter length. However, due to linear properties of the filters, as well as the assumed detector linearity, this ringing is entirely deterministic, and can therefore be modeled. Specifically, a filtered data segment containing $N$ overlapped pulses can be modeled as the sum of the appropriately scaled and shifted filtered pulse templates. I.e.

\begin{equation}
\label{eq:model}
M(t) = \sum_{i=1}^{N} a_{i} P_{i}(t - t_{i})
\end{equation}

\noindent where each filtered pulse template $P_{i}$ has been interpolated to the appropriate sub-sample arrival time based on $t_{i}$. As illustrated in Fig.~\ref{fig:filDat}, the value of the filtered data at each pulse arrival time is no longer determined by the amplitude and arrival time of a single pulse, but instead depends on the amplitude and arrival time of every pulse within the template length.

The objective is to solve for the amplitudes $a_{i}$ and arrival times $t_{i}$ that predict the observed values in each of the two filtered data segments at each pulse arrival time.  In practice, the solution is found through fitting, minimizing the quantity

\begin{equation}
\label{eq:chisqure}
\chi^{2} = \sum_{i=1}^{N} [D_{A}(t_i) - M_{A}(t_{i})]^2 + [D_{T}(t_i) - M_{T}(t_{i})]^2
\end{equation}

\noindent where $D_{A}$ and $M_{A}$ correspond to the amplitude-filtered data and model, respectively, while $D_{T}$ and $M_{T}$ correspond to the arrival time-filtered equivalents.  Models $M_{A}$ and $M_{T}$ each take the form of Eq.~\ref{eq:model}, incorporating the correspondingly filtered pulse templates for $P_{i}$ in each term. Note that, while each model can be evaluated at all times, only the pulse arrival times $t_{i}$ are considered in the fit. Therefore, for N pulses, there are 2N observations (the pulse arrival times in each of the two filtered data streams) and 2N parameters (the amplitude and arrival time for each pulse). This ensures that there will always be an exact solution, even in the presence of noise (though the noise will affect the accuracy of the derived solution).

Until now, we have made no mention of how the number of pulses in each fit segment is determined. In our technique, we opt to detect pulses in the amplitude-filtered data, since it provides the highest signal to noise by construction. However, since the ringing caused by the largest pulses can be comparable in amplitude to smaller pulses, pulses must be detected iteratively, starting with the largest.

In the initial pass, the peaks of all excursions larger than the maximum expected ringing are identified as pulses in the amplitude-filtered data segment, and initial guesses for their amplitudes and arrival times are made from the value and time of these peaks, respectively.  These guesses can used in Eq.~\ref{eq:model} to construct models for both the amplitude- and arrival time-filtered data segments, which can be further refined by minimizing Eq.~\ref{eq:chisqure} through the fitting procedure. The amplitudes and arrival times found for these pulses may be slightly distorted by as yet undetected smaller pulses, but subtracting this solution removes their associated ringing to a good approximation and allows the pulse detection threshold to be lowered on the next iteration. With each iteration, newly detected pulses are added to the models, and a new self-consistent solution for amplitudes and arrival times is found through fitting. This process continues until all pulses above the detector noise level are detected and included in the fit (typically achieved after three iterations).

Note that, since we are working with the amplitude-filtered data, the noise level is intrinsically related to the optimally filtered energy resolution. Based on our original formulation in Eq.~\ref{eq:estim} and assuming that the noise is uncorrelated at different frequencies, the mean-square fluctuations in our estimation of energy is given by

\begin{equation}
\label{eq:rsltn}
\sigma_{E}^{2} = \sum_{j=1}^{\infty} (w_{j} n_{j})^{2}
\end{equation}

\noindent
This is equivalent, however, to the mean-square fluctuations in the amplitude-filtered output at \textit{all} times, not just at the pulse arrival times. In practice, this means that the \mbox{$5\sigma$} lower threshold for pulse detection  normally considered adequate for a negligible false trigger rate is only about twice the optimally filtered FWHM (where $\text{FWHM}=2\sqrt{2\ln(2)}\sigma_{E}\sim2.35\sigma_{E}$). This is considerably better than the triggering techniques using simple filters that are typically employed in COF. The very low threshold not only enables detecting lower energy pulses, but also reduces the impact of undetected pulses corrupting the fit.

Pulses that arrive within a filter-length of the end of each segment may be affected by ringing of pulses that arrive in the following segment but are yet undetected. For this reason, consecutively analyzed segments are chosen to overlap by at least the filter length. Pulses whose templates are completely contained in one segment are subtracted from the overlapped region in the following segment. The remaining pulses, which may be affected by pulses in the following segment and vice-versa, are passed to the following segment to be refit with this new information. This process ensures that the final fit of every pulse in the data stream includes the information of every other pulse that arrives within a filter-length before or after it.

\subsection{Gain Correction}\label{sec:gain}
\begin{figure}
\centering{}
\includegraphics[width=3.37in]{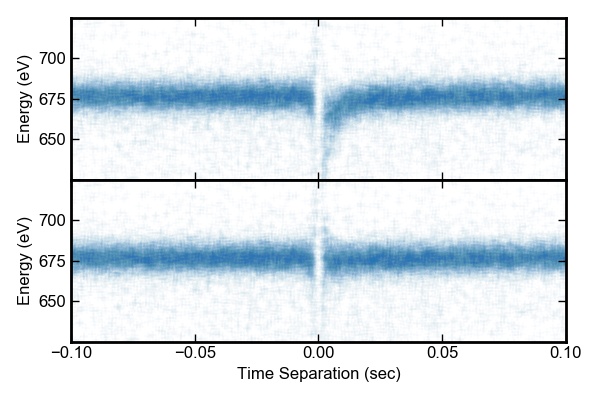}
\caption[Pileup Gain Correction]{Scatter plot of recovered pulse energy versus time separation for pulses created by 677~eV photons (F~K$\alpha$). Each point corresponds to a single 677~eV pulse that overlaps with another pulse at $t=0$, ranging in energy from 277--3590~eV (with the majority being $>3$~keV). Points plotted at negative times arrive prior to the other pulse, positive times after. Top: Pulses that arrive shortly after another pulse (within $\sim20$~ms, or a couple decay times) experience low detector gain due to the pixel still cooling from the preceding pulse. Bottom: Same pulses, after correcting for this effect. Spectral resolution does not degrade until pulses are separated by as little as a few rise times, at which point the sum of pulses cannot be distinguished from a single pulse.}
\label{fig:scatter} 
\end{figure}

The best-fit model parameters provide a ``raw'' pulse amplitude that needs to be converted to the photon energy. Since resolving powers ($E/\Delta E$) of these detectors can exceed several thousand, even small non-linearities that have negligible affect on pulse shape or noise must still be taken into account in this conversion. Typically, detector non-linearity is primarily due to non-linear thermometer response, and to a lesser extent to increased pixel heat capacity at higher temperatures. As a result, the pulse amplitude depends not only on the photon energy, but also on the pixel temperature at the time of pulse arrival. For pulses that arrive a sufficiently long time after the preceding pulse, the pixel is in thermal equilibrium at a temperature precisely regulated by the refrigerator. However, for pulses more closely spaced in time, the pixel is still cooling from the first pulse at the arrival time of the second. The second pulse will therefore begin at a higher pixel temperature, resulting in reduced detector response. Nevertheless, in the limit that the pulse shape and noise properties are not significantly impacted by this effect, the energy of the second pulse can still be recovered without loss of resolution, given its raw amplitude and the detector temperature at the time of arrival. As shown in Fig.~\ref{fig:scatter} this correction works well even when closely spaced pulses have very different energies. In practice, the function used to convert amplitude and temperature to photon energy can be determined from an adequate amount of calibration data, or else calculated from a detector model. The resulting function can be stored as a 2D table, which can later be efficiently interpolated.

\subsection{Addressing Changing Pulse Shape}\label{sec:shape}
For data sets containing pulses spanning a broad energy range (as is the case for many microcalorimeter applications), the shape of the largest pulses may be affected by detector non-linearity. For segments containing such large pulses, simply scaling a filtered pulse template determined from lower energy pulses will no longer provide an accurate model of the filtered data.  As a result, the fit solution for any other pulses arriving within the filter length of such a large pulse may be affected. To address this, we create a dictionary of pulse templates which can be selected or interpolated based on the pulse amplitude. In other words, the templates $P_{i}$ used in Eq.~\ref{eq:model} are made to be a function of both $t_{i}$ and $a_{i}$. These amplitude-dependent pulse shapes can be determined empirically with calibration data, or else modeled with knowledge of the particular detector. We note that, as these shapes only depend on the amplitude of a single pulse, they do not account for shape changes that result due to overlap. More elaborate algorithms could in principle account for this effect as well, but have not been implemented in this work. Moreover, in the limit of extreme non-linearity, the assumption of stationary noise breaks down and altogether different filtering techniques are needed (e.g. Ref.~\onlinecite{fixsen_optimal_2002}).

\subsection{Livetime Determination}
The OLPF technique can yield a live-time efficiency of nearly $100\%$ up to high count rates. The only irreducible source of dead-time arises from pulses with time separations comparable to the rise time of the detector ($\sim1.5$~ms for XQC). As the time separation of two pulses approaches this limit, the combined shape of the pulses begins to approximate the shape of a single pulse, and only the sum of pulse amplitudes is constrained. Indeed, in the limit of zero separation, the event will be indistinguishable from a single pulse with the total energy of the two pulses. However, the fraction of such pulses is very small for count rates much less than the inverse of the detector rise time.

Another source of dead-time encountered in practice is the inability to fit and subtract features in the convolved data streams with unexpected shapes. Such features include thermal and electrical cross-talk with other pixels, direct photon hits to the thermometer, and large pulses that saturate the electronics. Without templates for fitting these events, they cannot be included properly in the model, resulting in residuals that impact the solution for amplitudes and arrival times of good pulses. While it is possible to construct templates for such non-pulse features, in the data sets examined here it was not worth the small improvement in live-time compared to simply skipping over affected segments. To minimize the live-time cost associated with these non-pulse features, affected segments were masked from the data stream prior to convolution, to avoid ``smearing out'' the effects of the features over the filter length. 

\section{Computational Requirements}
The purpose of this section is only to provide a general sense of the computational requirements for the OLPF technique as it is currently implemented. The process is coded entirely in Python, and is executed on a modestly equipped desktop computer (4-core, 3.2~GHz processor; 12~GB of RAM). While some effort has been made to improve the efficiency of the algorithm, we have not conducted a detailed optimization. In practice, the minimization described in Sec.~\ref{sec:fit} is done with MINPACK's Levenberg-Marquardt non-linear least squares algorithm (Ref.~\onlinecite{more_user_1980}), since the estimation of arrival-time lacks an analytic solution (i.e. the solution cannot be found by matrix inversion). This is currently the most computationally intensive step of this process, accounting for $>60\%$ computation time. Therefore, the computation time depends more on the number of pulses than it does on the length of the data record. A single 3.2~GHz processor core can process $\sim10$~pulses/second using the current algorithm. This processing requirement puts the algorithm out of reach for real-time on-board analysis of a large array on a spacecraft.

\section{Performance Tests}\label{sec:test}
To evaluate the OLPF technique, we have compared its performance to COF using two distinct data sets. First we consider XQC data, on which the algorithm was developed and which is representative of many microcalorimeter applications. In particular, it contains a high relative rate of photons spanning a large range in energy, though the response is still approximately linear over this range. Furthermore, the noise spectra are not entirely smooth, so the spectral resolution is more sensitive to the choice of filter length (i.e. compared to smooth noise spectra, which primarily suffer from signal lost in the zeroth frequency bin alone). The X-rays in this data set are generated by two sources: a $^{41}$Ca source and an alpha-excited multi-target fluorescent source. Together, these sources produce K-shell emission lines of C, O, F, Al, Si, and K, spanning 277--3590~eV in energy. The combined photon count rate is 1.8 counts/second/pixel, with slightly more than half the photons originating from the $^{41}$Ca source ($>3$~keV).

\begin{figure}
\subfloat[\label{fig:histo}Comparison of Total Counts]{
\includegraphics[width=1.67in]{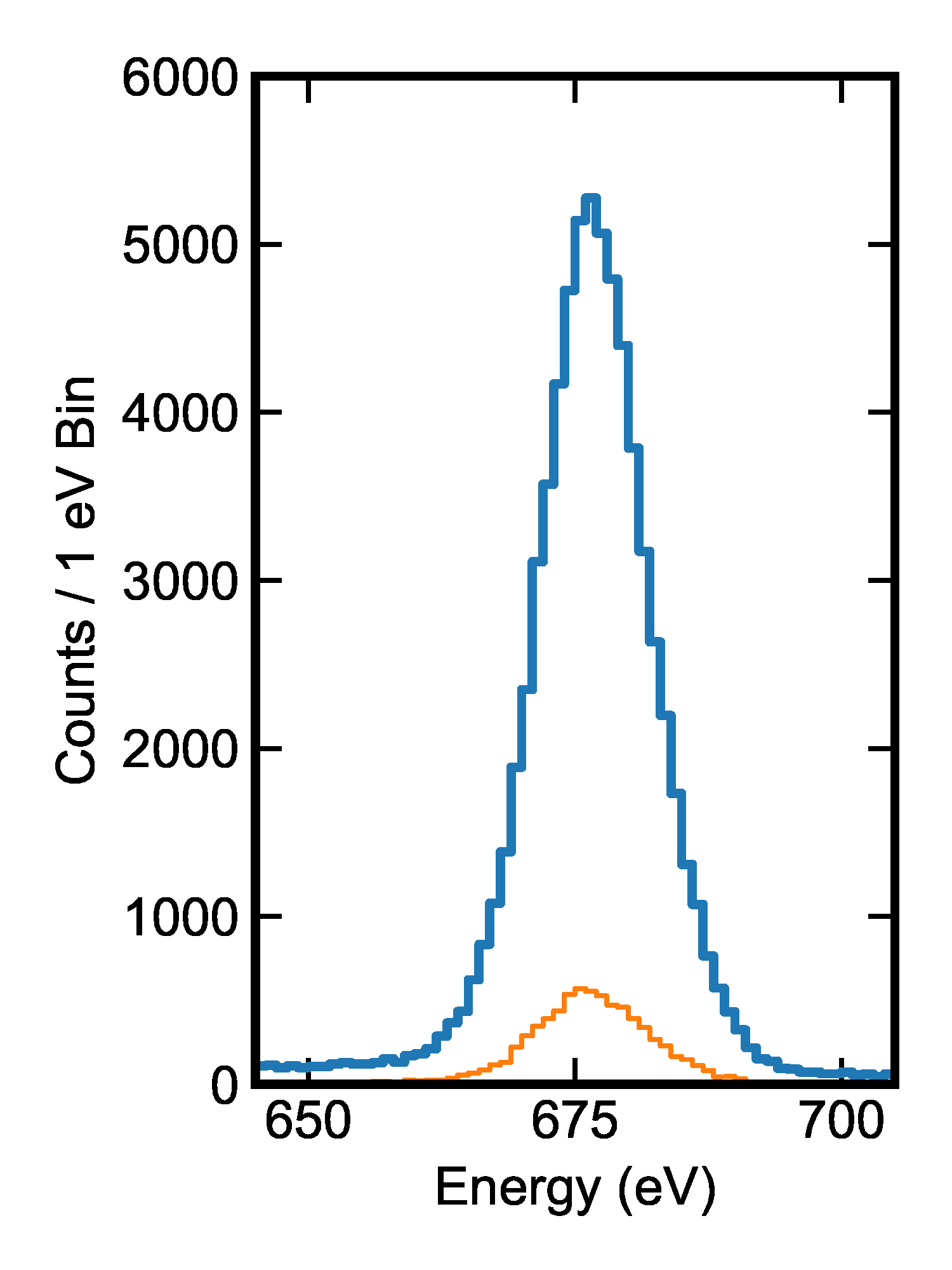}}
\subfloat[\label{fig:histoNorm}Comparison of Normalized Counts]{
\includegraphics[width=1.67in]{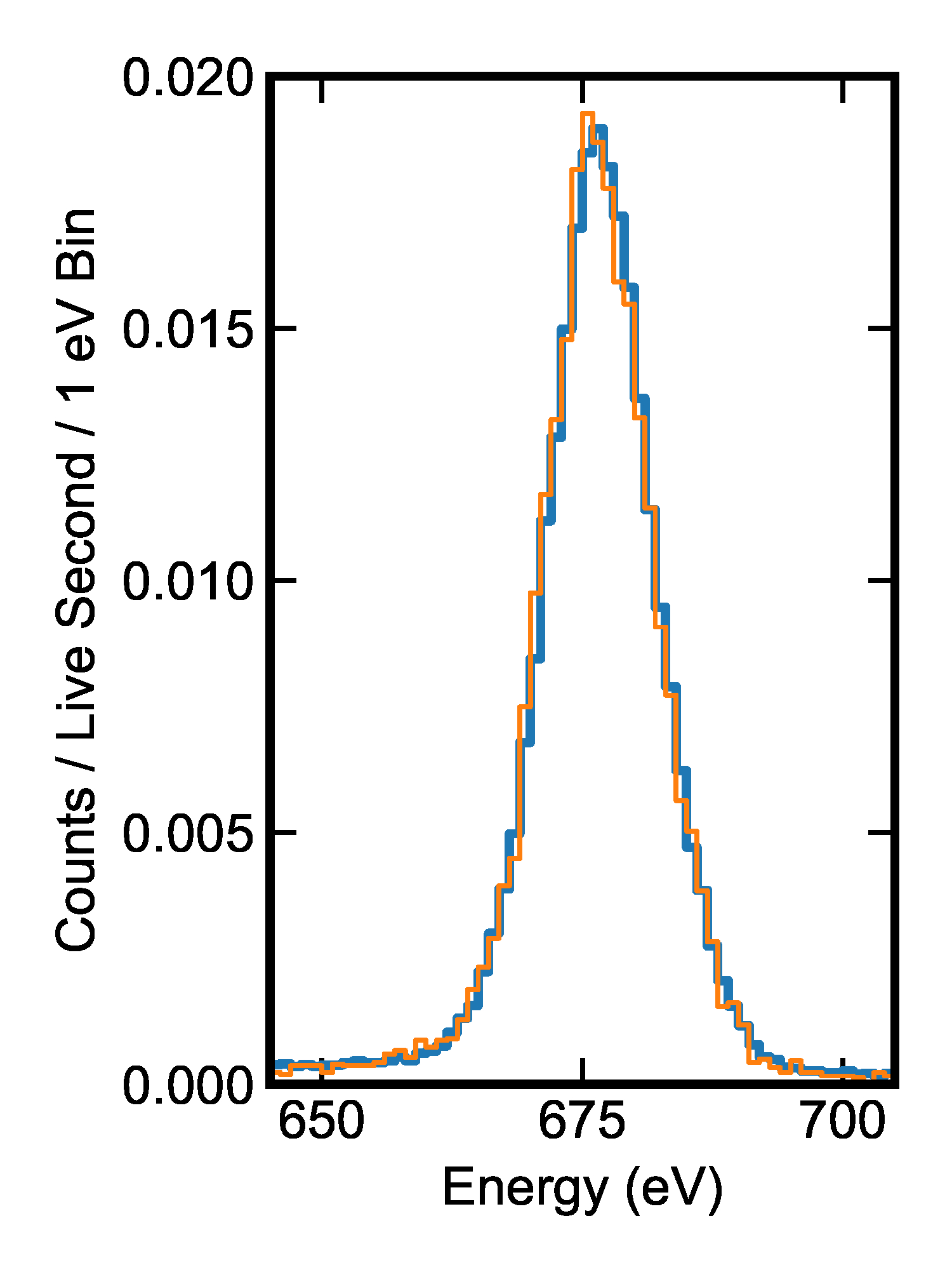}}
\caption[XQC Overlapped Pulse Resolution]{ XQC spectra of F K$\alpha$ line at 677~eV processed with the OLPF technique (heavy blue line) compared to COF of the same data set using a 200~ms filter (light orange line). (a) Histograms of total counts illustrate the $>8\times$ improvement in processing live-time with the OLPF technique. (b) Histograms normalized to live-time illustrate identical spectral resolution. Solid state effects broaden and distort the intrinsic line shape, making it difficult to accurately determine the instrumental resolution. Nevertheless, in both cases near-baseline performance of $\sim$7~eV~FWHM is achieved.}
\label{fig:histos}
\end{figure}

Fig.~\ref{fig:histos} depicts the spectra of the F K$\alpha$ line from the two different processing techniques. At this count rate, the processing live-time of COF is only $12\%$ when using the 200~ms filter needed for near-optimum spectral resolution. In comparison, the OLPF technique presented here achieves $98\%$ live-time while achieving identical spectral resolution to COF (the $2\%$ dead-time is almost entirely due to background events saturating the electronics). Under these operating conditions, the OLPF performs near its theoretical limit, recovering pulses separated by as little as the detector rise time without loss of resolution.

Next, we evaluate the performance of the OLPF technique using the data set presented in Ref.~\onlinecite{lee_high_2014}, which features $\sim$100~counts/second/pixel of 6~keV photons produced by a single $^{55}$Fe source. The small-pitch TES detector used here was optimized for high count rates, achieving an effective decay time $\tau\sim$200~$\mu$s through a combination of high thermal conductivity and low heat capacity. As noted in Ref.~\onlinecite{lee_high_2014}, the superconducting weak-link effect leads to better linearity than one may naively expect for devices with such small heat capacity. Nevertheless, at 6~keV this detector is still significantly more non-linear than typical TES or silicon thermistor microcalorimeters designed for observations at this energy. Isolated 6~keV pulses are already $\sim 30\%$ non-linear, and overlapped pulse pairs can easily drive the detector near saturation. Under these rather extreme conditions, the assumptions underlying both COF and OLPF no longer hold true. 

The analysis presented by Ref.~\onlinecite{lee_high_2014} utilized a graded optimal filtering technique adapted from that developed for the SXS instrument on Hitomi (Ref.~\onlinecite{ishisaki_-flight_2018}). This technique achieves almost $100\%$ processing live-time by using three filters of different lengths depending on pulse separation. ``High-res'' events are those which can be processed with the longest filter and therefore the highest spectral resolution, ``Mid-res'' events are those processed with a filter half the length, and ``Low-res'' events are those with only a boxcar smoothed estimation of pulse height.

While both the graded optimal filtering and OLPF techniques achieve essentially $100\%$ processing live-time for this data set, in Table~\ref{ta:tes} we compare the FWHM resolution of pulses in each event grade.  In the limit of long pulse separation (i.e. ``High-Res'' events), the two analyses reduce to the same conventional technique, so it is not surprising that the resolution does not depend on the choice of algorithm. For ``Mid-'' and ``Low-res'' graded events, both analyses exhibit degraded resolution to a comparable degree, with the OLPF technique delivering only slightly better performance.

\begin{table} 
\caption{Performance comparison of OLPF and COF techniques with small-pitch TES data. Note that all pulses/event grades are processed simultaneously with the OLPF technique.  Post processing, pulses are sorted into event grades according to inter-pulse separation for the sake of comparison with COF. Each technique achieves a total processing live-time of $>99\%$. COF resolution values reproduced from S. J. Lee et al., Journal of Low Temperature Physics \textbf{176,} 597 (2014); licensed under a Creative Commons Attribution (CC BY) license.}
\label{ta:tes}
\begin{ruledtabular}
\begin{tabular}{lccc}
 & & COF & OLPF \\
 & Percentage of & Resolution & Resolution \\
Event Grade & Total Counts & (eV FWHM) & (eV FWHM) \\
\hline
``High-res'' & 69.0 & $1.95 \pm 0.04$ & $1.95 \pm 0.11$ \\
``Mid-res'' & 10.1 & $2.25 \pm 0.11$ & $2.11 \pm 0.32$ \\
``Low-res'' & 20.4 & $3.78 \pm 0.08$ & $3.46 \pm 0.16$ \\
\end{tabular}
\end{ruledtabular}
\end{table}

While the shorter filter employed for the ``Mid-res'' pulses does contribute to the degradation in resolution for the graded COF technique, pulse shape changes due to detector non-linearity also contribute, limiting the performance of the OLPF technique as well. As noted in Sections~\ref{sec:gain} and \ref{sec:shape}, the gain correction currently implemented for overlapped pulses only corrects for linear changes in pulse amplitude, and does not account for changes in pulse shape due to overlap. For the ''Mid-'' and ''Low-res'' graded events in the this data set, the magnitude of the required gain correction is tens to hundreds of eV (in contrast to the XQC data, where this correction was typically $<$20~eV). As a result, even shape changes at the sub-percent level can impact the achieved resolution on the order of 1~eV. Much better performance would be expected for data dominated by lower-energy photons, which is astrophysically more realistic and would only rarely have overlaps of two large pulses that extend into the saturating regime. Unfortunately, such a data set with sufficiently high rates was not available.

\section{Conclusions}
We have presented an Overlapped Pulse Fitting technique for processing microcalorimeter data, which is capable of recovering pulses separated by as little as the rise time of the detector without loss of spectral resolution compared to COF. The improvement in throughput and/or resolution is greatest for data sets that approximately satisfy the assumptions of linear response and stationary noise. Even in instances where the highest energy pulses violate these assumptions, a significant fraction of the remaining pulses may still be recovered.

As microcalorimeters have gained in popularity in a variety of contexts, this technique may have wide-reaching applications across many scientific disciplines---particularly on ground-based experiments or space-based experiments with the option of offline processing. It may be possible to reduce computation time modestly with optimization of the existing algorithm and its coding. However, it is unlikely that the major reductions necessary for implementation on spacecraft will be possible without making some simplifying approximations to the overlapped pulse solution.

\begin{acknowledgments}
We wish to thank Scott Porter and the other authors of Ref.~\onlinecite{lee_high_2014} for sharing their data for this analysis. This work was supported by NASA grant NNX16AM31G.
\end{acknowledgments}

\section*{Data Availability}
The XQC data presented here are available from the corresponding author upon reasonable request. Restrictions apply to the small-pitch TES data, which were used with permission from the authors of Ref.~\onlinecite{lee_high_2014}. These data are also available from the authors upon reasonable request and with the permission of the authors of Ref.~\onlinecite{lee_high_2014}.

\bibliography{overlapped_pulses}

\end{document}